\documentclass[11pt]{article}
\usepackage{jheppub}
\usepackage{bm} 
\usepackage{soul} 

\def\coeff#1#2{{\textstyle {\frac {#1}{#2}}}}
\def\p{{\bf p}}

\def\f{{\bf f}}

\def\C{{\cal C}}
\def\E{{\cal E}}
\def\J{{\cal J}}
\def\L{{\cal L}}
\def\N{{\cal N}}
\def\P{{\cal P}}
\def\Q{{\cal Q}}
\def\T{{\cal T}}

\def\pt{p_\perp}

\def\ce{\varepsilon}
\def\H{{\cal H}}
\def\g{{\bf g}}
\def\A{{\bf A}}

\newcommand{\pder}[2]{\frac{\partial #1}{\partial #2}}
\newcommand{\braket}[1]{\langle #1 \rangle}

\title{Causal first-order hydrodynamics from kinetic theory and holography}

\author{Raphael E. Hoult and Pavel Kovtun}
\affiliation{Department of Physics \& Astronomy,  University of Victoria, \\ PO Box 1700 STN CSC, Victoria, BC,  V8W 2Y2, Canada}

\abstract{
We show how causal relativistic Navier-Stokes equations arise from the relativistic Boltzmann equation: the causality is preserved via a judicious choice of the zero modes of the collision operator. A completely analogous procedure may be used to extract causal hydrodynamics from the fluid-gravity correspondence: again, causality of the hydrodynamic equations is preserved by a suitable choice of zero modes of the corresponding differential operators in the bulk. We give examples of zero modes which give rise to causal hydrodynamic equations for non-conformal fluids with a conserved U(1) global symmetry current.
}

\begin{document}
\maketitle

\section{Introduction}
Relativistic viscous hydrodynamics is by now over eighty years old, starting with the classic works by Eckart~\cite{PhysRev.58.919} in 1940, and by Landau and Lifshitz in the second edition of their book~\cite{LL6} in 1953. Following the treatment of non-relativistic fluids, the classic theories of relativistic hydrodynamics introduced dissipative effects via terms which have derivatives of the standard hydrodynamic variables: fluid velocity $u^\alpha$, temperature $T$, and the chemical potential~$\mu$. Schematically, the energy-momentum tensor $T^{\alpha\beta}$ and the particle number current~$J^\alpha$ of the classic theories take the form of the following constitutive relations:
\begin{align}
\label{eq:TJ-1}
  T^{\alpha\beta}, J^\alpha = O(T,u,\mu) + O(\partial T, \partial u, \partial \mu)\,,
\end{align}
where the first term in the right-hand side corresponds to perfect fluids, and the second term encodes dissipative corrections due to viscosity and heat conductivity. The hydrodynamic equations are the conservation laws 
\begin{align}
\label{eq:cons}
  \partial_\alpha T^{\alpha\beta}=0\,,\ \ \ \ \partial_\alpha J^\alpha = 0\,,
\end{align}
which give partial differential equations for $T$, $u$, and $\mu$. The classic theories are often called ``first-order'' theories, because the above $T^{\alpha\beta}$ and $J^\alpha$ contain only up to one derivative of the hydrodynamic variables. It was understood soon after that the classic hydrodynamic theories suffer from violations of causality. From a mathematical point of view, the hydrodynamic equations of the classic theories are not hyperbolic.  Related to that, the classic theories predict that the uniform equilibrium state of a non-gravitating fluid in flat space is unstable. See Refs.~\cite{Hiscock:1985zz} and \cite{Hiscock:1987zz} for an extensive discussion of instability and acausality in the classic theories. 

The most popular proposal to remedy the violations of causality in the classic formulations are the M\"uller-Israel-Stewart (MIS) theories \cite{Muller:1967, Israel:1976tn,Israel-Stewart}. Schematically, MIS theories introduce extra tensor variables collectively denoted by $\Pi$, and posit that
\begin{align}
\label{eq:MIS-1}
  & T^{\alpha\beta}, J^\alpha = O(T,u,\mu) + \Pi\,,\\
\label{eq:MIS-2}
  & \tau\, u^\lambda \partial_\lambda \Pi = O(\Pi, \partial T, \partial u, \partial \mu)\,,
\end{align}
where $\tau$ denote phenomenological ``relaxation times'' introduced in the MIS theory. The terms  $O(\partial T, \partial u, \partial\mu)$  in the right-hand side of \eqref{eq:MIS-2} are inherited from the classic first-order theories. Combined with the standard conservation laws $\partial_\alpha T^{\alpha\beta}=0$ and $\partial_\alpha J^\alpha = 0$, one finds partial differential equations for $\Pi$, $u$, $T$, and $\mu$. The MIS equations are hyperbolic provided the parameters and the variables of the theory satisfy certain constraints, though demonstrating hyperbolicity in the full non-linear theory is tricky, and was only accomplished recently~\cite{Bemfica:2020xym}.

The introduction of numerous extra variables $\Pi$ in the MIS theory seems like a drastic modification of the classic hydrodynamic formulation~\eqref{eq:TJ-1}, and it is. Moreover, there is no need to introduce extra variables $\Pi$ to maintain causality: the classic theories can be made consistent with causality with only minimal modification, keeping the first-order structure~\eqref{eq:TJ-1} intact. The reason is that the classic theories~\cite{PhysRev.58.919, LL6} do not include all possible one-derivative terms in the right-hand side of the constitutive relations~\eqref{eq:TJ-1}. The choice of which one-derivative terms are included in eq.~\eqref{eq:TJ-1} can be traced to the convention for the out-of-equilibrium definitions of $T$, $u^\alpha$, and $\mu$. This convention is often called a choice of ``frame'', thus one speaks of ``Eckart frame'', ``Landau-Lifshitz frame'', etc. One can consider a most general ``frame'', which in practice amounts to including all possible one-derivative terms in the right-hand side of eq.~\eqref{eq:TJ-1}. One can choose ``frames'' in which the hydrodynamic equations~\eqref{eq:cons} are hyperbolic, the equilibrium state is stable, and the entropy production is positive within the domain of validity of the derivative expansion. Such ``good frames'' are the essence of the BDNK approach~\cite{Bemfica:2017wps, Kovtun:2019hdm, Bemfica:2019knx, Hoult:2020eho, Bemfica:2020zjp}. Formally, such theories are first-order theories because the constitutive relations still have the form~\eqref{eq:TJ-1}. Thus, if one's goal is to have a minimal causal theory of relativistic hydrodynamics, the MIS theory is overkill, and the more conservative first-order BDNK formulation is sufficient. It is straightforward to combine the two approaches by including all possible $O(\partial T, \partial u, \partial\mu)$ terms in the right-hand side of eq.~\eqref{eq:MIS-2}, see~\cite{Noronha:2021syv}. For a numerical exploration of causal first-order hydrodynamics see~\cite{Pandya:2021ief}.

The purpose of this paper is to show how the equations of causal first-order hydrodynamics arise from relativistic kinetic theory. Kinetic theory is not required for hydrodynamics: a fluid can happily flow and obey hydrodynamic constitutive relations on macroscopic distance and time scales even if the fluid is not composed of quasi-particles. Perhaps the most extreme examples come from fluids comprised of strongly interacting quantum fields in non-abelian gauge theories, found in the holographic correspondence~\cite{Aharony:1999ti}. Their hydrodynamic flows have been studied extensively, see e.g.~\cite{Chesler:2013lia}. In these examples, strong quantum fluctuations completely wash out the very notion of a quasi-particle, rendering kinetic theory inapplicable. On the other hand, for those fluids which {\em are} comprised of well-defined quasi-particles, the Boltzmann equation provides a useful simple model of macroscopic dynamics. 

The derivation of first-order causal hydrodynamics from kinetic theory has been addressed previously in refs.~\cite{Bemfica:2017wps, Bemfica:2019knx}. Our treatment here differs in several respects. We do not ignore the conserved particle number current, and we do not impose any ``matching conditions'' on the higher moments of the distribution function. We show how first-order causal hydrodynamics arises from the simplest derivative expansion (the Hilbert expansion) of the solutions to the Boltzmann equation, and that ``good frames'' arise simply as a ``good'' choice of the free functions which parametrize the zero modes of the linearized collision operator in real space. 

The method by which hydrodynamics is extracted from classical gravity in asymptotically anti-de Sitter spacetimes via the fluid/gravity correspondence \cite{Bhattacharyya:2008jc} is analogous in many ways to the Hilbert expansion in kinetic theory. The methods used to extract causal hydrodynamics from kinetic theory apply equally well to holography, allowing the construction of causal hydrodynamic theories with holographic duals. 

The structure of the paper is as follows. In section \ref{sec::BE}, we show how causal first-order hydrodynamics arises from the Hilbert expansion in relativistic kinetic theory. In section \ref{sec::fg-correspond}, we show how an analogous procedure leads to causal hydrodynamics via the fluid/gravity correspondence. In appendices~\ref{ap::concrete} and \ref{sec:realspace} we discuss examples of causal hydrodynamic frames, illustrating a causal choice of the zero modes. Finally, in appendix \ref{ap::matching}, we comment on the ``matching conditions" on the higher moments of the distribution function.

\section{Boltzmann equation}\label{sec::BE}
\noindent
Relativistic kinetic theory is an established subject~\cite{deGroot, Cercignani-Kremer}. The fundamental object of the theory is the one-particle distribution function $f(x,p)$, where both the particle's location $x$ and momentum $p$ are four-vectors. The particles are on-shell, with $p^0\equiv(\p^2 +m^2)^{1/2}$. The distribution function counts the particles, and is normalized so that the number density is 
\begin{align}
  n(x) = \int\!\! \frac{d^3p}{(2\pi)^3}\; f(x,p)\,.
\end{align}
The particle number current is the covariant version of the above:
\begin{align}
\label{eq:JJ}
  J^\mu = \int_p p^\mu f(x,p)\,,
\end{align}
where $\int_p \dots \equiv \int d^3p/[(2\pi)^3 p^0] \dots$ denotes the Lorentz-invariant integration measure. Similarly, the energy-momentum tensor is
\begin{align}
\label{eq:TT}
  T^{\mu\nu} = \int_p p^\mu p^\nu f(x,p)\,.
\end{align}

\subsubsection*{Conservation laws}
The Boltzmann equation is the evolution equation for the distribution function,
\begin{align}
\label{eq:BE-1}
  p^\mu \partial_\mu f(x,p) = \C[f]\,,
\end{align}
where the right-hand side is the collision term which is an integral operator (in momentum) that acts on $f_p \equiv f(x,p)$, and is at least quadratic in $f$. The details of $\C$ depend on the interactions and on the statistics of the particles, and we will assume that the inter-particle interactions conserve energy, momentum, and particle number. A simple form to keep in mind is 2-to-2 elastic collisions%
\footnote{Upper sign is for bosons, lower is for fermions.}
\begin{align}
\label{eq:Cf}
  \C[f] = \coeff12 \int_{p_1, p_2, p_3} W(p,p_1 | p_2,p_3) \left[ f_{p_2} f_{p_3} (1\pm f_p) (1\pm f_{p_1}) -  f_{p} f_{p_1} (1\pm f_{p_2}) (1\pm f_{p_3}) \right] \,.
\end{align}
The transition rates obey $W(p, p_1 | p_2, p_3) = W(p_2, p_3 | p, p_1) = W(p_1, p | p_2, p_3) = W(p, p_1 | p_3, p_2)$, and are proportional to $\delta(p+p_1-p_2-p_3)$.
It then follows that
\begin{align}
\label{eq:C-cons}
  \int_p [a(x) + b_\mu(x) p^\mu] \, \C[f_p] = 0\,,
\end{align}
with arbitrary functions $a(x)$, $b_\mu(x)$. In particular, $\int_p \C[f_p] = 0$ and $\int_p p^\mu \C[f_p] = 0$. Together with the Boltzmann equation \eqref{eq:BE-1}, these immediately imply 
\begin{align}
\label{eq:JT-cons}
  \partial_\mu J^\mu = 0\,,\ \ \ \ \partial_\mu T^{\mu\nu} = 0\,.
\end{align}
Alternatively, the conservation laws \eqref{eq:JT-cons} can be viewed as the zeroth and first ``moments'' of the Boltzmann equation, i.e.\ they follow by applying $\int_p \dots$ and $\int_p p^\nu \dots$ to eq.~\eqref{eq:BE-1}, and are true for any $f(x,p)$. The property \eqref{eq:C-cons} of the collision operator is a manifestation of the conservation of energy, momentum, and particle number. Approximate forms of the collision operator which violate eq.~\eqref{eq:C-cons} (such as the relaxation-time approximation) are in general inconsistent with the conservation laws \eqref{eq:JT-cons} which form the basis of hydrodynamics.%
\footnote{
     A way to fix the relativistic relaxation-time approximation was discussed recently in ref.~\cite{Rocha:2021zcw}.
}

\subsubsection*{Equilibrium}
Consider distribution functions of the form
\begin{align}
\label{eq:ff}
  \f(x,p) = [ e^{-{\bm\beta}_\mu(x) p^\mu - {\bm\alpha}(x)} \mp 1]^{-1}\,,
\end{align}
with arbitrary ${\bm\beta}_\mu(x)$, ${\bm\alpha}(x)$. These are local Bose/Fermi distributions for a fluid at temperature ${\bm T} = 1/\sqrt{-{\bm\beta}^2}$, with velocity ${\bm u}^\alpha = {\bm\beta}^\alpha/\sqrt{-{\bm\beta}^2}$, and chemical potential ${\bm\mu} = {\bm\alpha}/\sqrt{-{\bm\beta}^2}$. 
The collision integral is such that it satisfies $\C[\f] = 0$. If we want the function (\ref{eq:ff}) to actually solve the Boltzmann equation, we need $p^\mu p^\nu \partial_\mu {\bm\beta}_\nu + p^\mu \partial_\mu {\bm\alpha} = 0$, hence the functions ${\bm\beta}_\mu(x)$ and ${\bm\alpha}(x)$ must satisfy
\begin{align}
\label{eq:ba-eq}
  \partial_\mu {\bm\beta}_\nu(x) + \partial_\nu {\bm\beta}_\mu(x) = 0\,,\ \ \ \  \partial_\mu {\bm\alpha}(x) = 0\,.
\end{align}
Generalized to curved space, the first equation would say that ${\bm\beta}_\mu$ is a Killing vector.

\subsection*{Derivative expansion in hydrodynamics}
Let us forget about kinetic theory for a moment, and consider hydrodynamics per se, with the constitutive relations written in the derivative expansion:
\begin{align}
  & T^{\mu\nu} = T^{\mu\nu}_{0}[\beta] + \gamma T^{\mu\nu}_{1}[\beta] + \gamma^2 T^{\mu\nu}_{2}[\beta] + \dots\,,
\end{align}
where $\gamma$ is a formal derivative-counting parameter, and $\beta^\mu=u^\mu/T$ are hydrodynamic variables: $T$ is the temperature, and $u^\mu$ is the fluid velocity. Truncating the above expansion at $O(\gamma^n)$ gives  $n$-th order hydrodynamics: $T^{\mu\nu}_{0} = O(\beta)$ is the perfect fluid, $T^{\mu\nu}_{1} = O(\partial\beta)$ contains the viscosity, $T^{\mu\nu}_{2} = O(\partial^2\beta, (\partial\beta)^2)$ etc. Let us look for the solutions using the same derivative expansion:
\begin{align}
 \beta^\mu(x) = \beta^\mu_{0}(x) + \gamma \beta^\mu_{1}(x) + \gamma^2 \beta^\mu_{2}(x) + \dots\,.
\end{align}
We expect $\beta_{n+1} = O(\partial \beta_n)$. Expanding the constitutive relations, we have 
\begin{align}
  T^{\mu\nu}_{n}[\beta] = T^{\mu\nu}_{n}[\beta_{0}] + \gamma\, T^{\mu\nu}_{n,1}[\beta_0, \beta_1] + \gamma^2\, T^{\mu\nu}_{n,2}[\beta_0, \beta_1, \beta_2] + \dots\,.
\end{align} 
The hydrodynamic variables $\beta$ are determined by solving $\partial_\mu T^{\mu\nu}[\beta]=0$, order by order in $\gamma$. At the leading order, the variables $\beta_0$ are determined by the perfect-fluid hydrodynamics: 
\begin{align}
\label{eq:b-0}
  \partial_\mu T^{\mu\nu}_{0}[\beta_{0}] = 0\,.
\end{align}
The first correction $\beta_1$ is then determined by 
\begin{align}
\label{eq:b-1}
  \partial_\mu \left( T^{\mu\nu}_{0,1}[\beta_0, \beta_1] + T^{\mu\nu}_{1}[\beta_0] \right) = 0\,.
\end{align}
At the next order,
\begin{align}
\label{eq:b-2}
  \partial_\mu \left( T^{\mu\nu}_{0,2}[\beta_0, \beta_1, \beta_2] + T^{\mu\nu}_{1,1}[\beta_0, \beta_1] + T^{\mu\nu}_{2}[\beta_0] \right) = 0\,,
\end{align}
determines the correction $\beta_2$, and the chain continues. The expansion \eqref{eq:b-0}, \eqref{eq:b-1}, \eqref{eq:b-2} etc.\ naturally arises from the derivative expansion in both kinetic theory, and in the fluid-gravity duality.  Note, however, that this is not how the hydrodynamic equations are normally solved for hydrodynamic variables. In practice, the hydrodynamic constitutive relations are given to the desired order in $\gamma$, and then the conservation equations are solved ``all at once" for $\beta^\mu$, as opposed to finding the order-by-order contributions $\beta_0, \beta_1$, etc. Such a procedure may lead to solutions which violate the small-derivative assumption of the expansion. The breakdown of the derivative expansion is a separate subject which we will not explore here.

\subsubsection*{Derivative expansion for the Boltzmann equation}
The distribution function \eqref{eq:ff} with arbitrary non-constant ${\bm\beta}_\mu(x)$ and ${\bm\alpha}(x)$ does not satisfy the Boltzmann equation. Approximate solutions to the Boltzmann equation may be constructed in the derivative expansion. To do so, we write the Boltzmann equation as 
\begin{align}
\label{eq:BE-1a}
  \varepsilon \, p^\mu \partial_\mu f(x,p) = \C[f]\,,
\end{align}
with an auxiliary parameter $\varepsilon$ (to be set to one at the end), and aim to find the solution as a power series in $\varepsilon$:
\begin{align}
  f(x,p) = \f(x,p) + \varepsilon f^{(1)}(x,p) + \varepsilon^2 f^{(2)}(x,p) + \dots\,.
\end{align}
This is sometimes called the Hilbert expansion~\cite{Cercignani-Kremer}. The energy-momentum tensor \eqref{eq:TT} and the current \eqref{eq:JJ} then take the form
\begin{align}
  & T^{\mu\nu} = T^{\mu\nu}_{(0)} + \varepsilon T^{\mu\nu}_{(1)} + \varepsilon^2 T^{\mu\nu}_{(2)} + \dots\,,\\
  & J^\mu = J^\mu_{(0)} + \varepsilon J^{\mu}_{(1)} + \varepsilon^2 J^{\mu}_{(2)} + \dots\,.
\end{align}
These expansions for $T^{\mu\nu}$ and $J^\mu$ are however not necessarily the derivative expansions of the hydrodynamic constitutive relations. In order to talk about the constitutive relations, we need the hydrodynamic variables $T$, $u^\lambda$, and $\mu$, or equivalently $\beta^\lambda = u^\lambda/T$ and $\alpha=\mu/T$. In kinetic theory, the hydrodynamic variables arise as arbitrary functions of $x$ (or ``integration constants'' in momentum space) in the solutions of the Boltzmann equation. The $x$-dependence of these functions is then fixed by the consistency conditions for the Boltzmann equation at each order in the expansion. These consistency conditions are exactly the hydrodynamic conservation laws. Each order in the $\varepsilon$-expansion generates its own arbitrary functions, namely
\begin{align}
\label{eq:bb-expansion}
  & \beta^\mu(x) = \beta^\mu_{(0)}(x) + \varepsilon \beta^\mu_{(1)}(x) + \varepsilon^2 \beta^\mu_{(2)}(x) + \dots\,,\\
\label{eq:aa-expansion}
  & \alpha(x) = \alpha_{(0)}(x) + \varepsilon \alpha_{(1)}(x) + \varepsilon^2 \alpha_{(2)}(x) + \dots\,,
\end{align}
where the leading-order hydrodynamic variables $\beta^\mu_{(0)} = {\bm\beta}^\mu$ and $\alpha_{(0)} = {\bm\alpha}$ are the free functions that appear in the equilibrium distribution (\ref{eq:ff}), and the corrections $\beta^\mu_{(n)}$ and $\alpha_{(n)}$ appear as undetermined functions in the solution for $f^{(n)}(x,p)$. Connecting to the earlier discussion of the derivative expansion in hydrodynamics, we expect to find in the Hilbert expansion 
\begin{subequations}
\begin{align}
\label{eq:TT-0}
  & T^{\mu\nu}_{(0)} = T^{\mu\nu}_{0}[\beta_{(0)}, \alpha_{(0)}]\,,\\
\label{eq:TT-1}
  & T^{\mu\nu}_{(1)} = T^{\mu\nu}_{0,1}[\beta_{(0)}, \alpha_{(0)}, \beta_{(1)}, \alpha_{(1)}] + T^{\mu\nu}_{1}[\beta_{(0)}, \alpha_{(0)}]\,,\\
\label{eq:TT-2}
  & T^{\mu\nu}_{(2)} = T^{\mu\nu}_{0,2}[\beta_{(0)}, \alpha_{(0)}, \beta_{(1)}, \alpha_{(1)}, \beta_{(2)}, \alpha_{(2)}] + T^{\mu\nu}_{1,1}[\beta_{(0)}, \alpha_{(0)}, \beta_{(1)}, \alpha_{(1)}] + T^{\mu\nu}_{2}[\beta_{(2)}, \alpha_{(2)}]\,,
\end{align}
\end{subequations}
etc., with analogous expressions for the current $J^\mu$. We expect the conservation equations to hold at each order in the expansion, 
\begin{align}
\label{eq:TJn-cons}
  \partial_\mu T^{\mu\nu}_{(n)}=0,\ \ \ \ \partial_\mu J^{\mu}_{(n)}=0\,,
\end{align}
This is indeed what happens.

\subsubsection*{First order: The equation}
At first order in the expansion we have
\begin{align}
\label{eq:BE-1b}
   \varepsilon \, p^\mu \partial_\mu  \f(x,p) = \C[\f + \varepsilon\, f^{(1)}]\,.
\end{align}
We expand the collision operator to linear order in $\varepsilon$. Denoting $\phi_p(x) \equiv f^{(1)}/\f(1\pm\f)$, the Boltzmann equation \eqref{eq:BE-1b} becomes
\begin{align}
\label{eq:BE-3}
  p^\mu \partial_\mu \f  = \f \, {\cal L}[\phi] \,,
\end{align}
where ${\cal L}$ is the linearized collision operator. Its explicit form depends on the details of the full collision operator $\C$, and in general one has
\begin{align}
\label{eq:L-zeromodes}
  {\cal L}[a(x) + b_\mu(x) p^\mu] = 0\,,
\end{align}
with arbitrary $a(x)$, $b_\mu(x)$. The existence of these zero modes is a consequence of $\C[\f] = 0$, reflecting the microscopic conservation laws of energy, momentum, and particle number. For 2-to-2 elastic collisions \eqref{eq:Cf}, the explicit form is
\begin{align}
\label{eq:Lf}
  {\cal L}[\phi] = \coeff12 \int_{p_1,p_2,p_3} W(p,p_1|p_2,p_3) \f_{p_1}(1\pm \f_{p_2}) (1\pm \f_{p_3}) \left( \phi_{p_2} + \phi_{p_3} - \phi_p - \phi_{p_1} \right) \,.
\end{align}

\subsubsection*{First order: The constraint}
Given two functions $g_p\equiv g(x,p)$, $h_p\equiv h(x,p)$, the linearized collision operator satisfies
\begin{align}
\label{eq:L-symmetry}
  \int_p \f_p\, g_p \, {\cal L}[h] = \int_p \f_p\, h_p \, {\cal L}[g]\,.
\end{align}
If we take $h_p=a(x) + b_\mu(x) p^\mu$ with arbitrary $a(x)$, $b_\mu(x)$, take $g_p=\phi_p$, and use the linearized Boltzmann equation \eqref{eq:BE-3}, we immediately find
\begin{align}
  0 = a(x)\, \partial_\mu {\bf J}^\mu + b_\mu(x)\, \partial_\lambda  {\bf T}^{\lambda\mu}\,.
\end{align}
In other words, at first order in the $\varepsilon$-expansion, the functions ${\bm\beta}_\mu(x)$ and ${\bm\alpha}(x)$ that appear in the local-equilibrium distribution function~\eqref{eq:ff} must obey
\begin{align}
\label{eq:JT-cons-0}
  \partial_\mu  {\bf J}^\mu = 0\,,\ \ \ \ \partial_\mu  {\bf T}^{\mu\nu} = 0\,.
\end{align}
These are the perfect-fluid conservation equations. The above $ {\bf T}^{\mu\nu}$ and ${\bf J}^\mu$ can be written as
\begin{align}
\label{eq:JT-0}
   {\bf T}^{\mu\nu} = {\bm\epsilon} {\bm u}^\mu {\bm u}^\nu +  {\bm p} {\bm\Delta}^{\mu\nu}\,,\ \ \ \ 
   {\bf J}^\mu =  {\bm n} {\bm u}^\mu\,,
\end{align}
where ${\bm\Delta}^{\mu\nu} \equiv g^{\mu\nu}+ {\bm u}^\mu {\bm u}^\nu$, and the coefficients are
\begin{align}
  &  {\bm n}({\bm T}, {\bm\alpha}) = -{\bm T} {\bm\beta}_\mu \int_p p^\mu \, \f_p \,,\\
  &  {\bm \epsilon}({\bm T}, {\bm\alpha}) = {\bm T}^2 {\bm\beta}_\mu {\bm\beta}_\nu \int_p p^\mu p^\nu \, \f_p \,,\\
  &  {\bm p}({\bm T}, {\bm\alpha}) = \coeff13 {\bm\Delta}_{\mu\nu} \int_p p^\mu p^\nu \, \f_p\,,
\end{align}
corresponding to the ideal-gas particle number density, energy density, and pressure. In the notation of eq.~\eqref{eq:TT-0}, ${\bf T}^{\mu\nu} = T^{\mu\nu}_0$, ${\bf J}^{\mu} = J^{\mu}_0$.
The conservation equations \eqref{eq:JT-cons-0} are
\begin{subequations}
\label{eq:JT-cons-1}
\begin{align}
  & n \,\partial{\cdot}u + (\partial n/\partial T) \dot T + (\partial n/\partial\alpha) \dot\alpha = 0 \,,\\
  & (\partial \epsilon/\partial T) \dot T + (\partial \epsilon/\partial\alpha) \dot\alpha + (\epsilon{+}p) \partial{\cdot} u = 0\,,\\
  & (\epsilon{+}p) \dot u_\mu + (\partial p/\partial T) \partial_\mu^\perp T + (\partial p/\partial\alpha) \partial_\mu^\perp \alpha = 0\,,
\end{align}
\end{subequations}
where all quantities are of order $O(\varepsilon^0)$. The dot stands for $u^\lambda \partial_\lambda$, and $\partial_\mu^\perp \equiv \Delta_{\mu\nu}\partial^\nu$. 
The vector conservation equation can be rewritten as $\dot u_\mu + \partial_\mu^\perp T/T + \frac{nT}{\epsilon+p} \partial_\mu^\perp \alpha = 0$.

Another way to arrive at eqs.~\eqref{eq:JT-cons-0} is to note that in the Boltzmann equation \eqref{eq:BE-3} the linearized collision operator ${\cal L}$ has zero modes, and therefore is not invertible. In general, the linear equation $H = {\cal L}[\phi]$ can only be solved for $\phi$ if the left-hand side $H$ is orthogonal to the zero-modes of the operator ${\cal L}$ in the right-hand side. For the linearized Boltzmann equation, the zero-modes are 1 and $p^\lambda$, and the consistency conditions amount to 
\begin{align}
  \int_p \f_p\, H = 0\,,\ \ \ \ \int_p \f_p\,  H\,  p^\lambda = 0\,,
\end{align}
with $H = p^\mu\partial_\mu \f_p/\f_p$. This again gives  eq.~\eqref{eq:JT-cons-0}. In other words, the equations of $0^{\rm th}$-order (perfect-fluid) hydrodynamics arise as constraint equations at $1^{\rm st}$-order in the expansion.

\subsubsection*{First order: Homogeneous solution}
At the first order in the expansion we have to solve eq.~\eqref{eq:BE-3} which we write as $H = {\cal L}[\phi]$, with $H = p^\mu\partial_\mu \f_p/\f_p$. The solution can be written as $\phi_p(x) = a(x) + b_\mu(x) p^\mu + \Phi(x,p)$ where $a(x)$ and $b_\mu(x)$ are arbitrary, and the inhomogeneous solution $\Phi$ satisfies $\Phi|_{H\to0}=0$. Remembering the definition $\phi_p = f^{(1)}/\f(1\pm\f)$, the distribution function to first order in $\varepsilon$ is 
\begin{align}
\label{eq:ff-22}
  f_p  
      = \f_p + \f_p (1\pm \f_p) \left( \varepsilon a + \varepsilon b_\mu p^\mu \right)  + \f_p (1\pm \f_p) \varepsilon \Phi + O(\varepsilon^2)\,.
\end{align}
From here, it is clear that the functions $b_\mu(x)$ and $a(x)$ can be understood as $O(\varepsilon)$ redefinitions of the functions $\beta_\mu(x)$ and $\alpha(x)$ which sit in $\f_p$. Indeed, for $\beta_\mu(x) = \beta_\mu^{(0)}(x) + \varepsilon \beta_\mu^{(1)}(x)$, $\alpha(x) = \alpha^{(0)}(x) + \varepsilon \alpha^{(1)}(x)$ we have in terms of $\f_p^{(0)} \equiv \f_p(\beta^{(0)}, \alpha^{(0)})$:
\begin{align}
\label{eq:ff-23}
  \f_p(\beta,\alpha) = \f_p^{(0)} +\f_p^{(0)} (1\pm \f_p^{(0)}) \left( \varepsilon \alpha^{(1)} +  \varepsilon \beta_\mu^{(1)} p^\mu \right) + O(\varepsilon^2) \,.
\end{align}
Alternatively, when we evaluate the energy-momentum tensor $T^{\mu\nu}$ and the current $J^\mu$ using the distribution function (\ref{eq:ff-22}), the only effect of the ``integration constants'' $b_\mu(x)$ and $a(x)$ is a linearized redefinition of ${\bm\beta}_\mu(x)$ and ${\bm\alpha}(x)$ in the perfect-fluid ${\bf T}^{\mu\nu}$ and ${\bf J}^\mu$. We thus identify the correction to the hydrodynamic variables in (\ref{eq:bb-expansion}) and (\ref{eq:aa-expansion}) as $\beta^\mu_{(1)} = b^\mu$, $\alpha_{(1)} = a$, keeping in mind that $b^\mu$ and $a$ are arbitrary, hence the fluid velocity, temperature, and the chemical potential at $O(\varepsilon)$ are intrinsically ambiguous quantities. Explicitly, the function (\ref{eq:ff-23}) leads to the shift of  ${\bm T} = T_{(0)}$, ${\bm\alpha} = \alpha_{(0)}$, and ${\bm u}^\mu = u^\mu_{(0)}$ in the perfect-fluid expressions (\ref{eq:JT-0}) by 
\begin{align}
  & T_{(0)} \to T_{(0)} + \varepsilon T_{(1)} = {\bm T} + \varepsilon {\bm T}^2 {\bm u}^\lambda b_\lambda \,,\\
  & \alpha_{(0)} \to \alpha_{(0)} + \varepsilon \alpha_{(1)} = {\bm\alpha} + \varepsilon a\,, \\
\label{eq:u-01}
  & u^\mu_{(0)} \to u^\mu_{(0)} + \varepsilon u^\mu_{(1)} = {\bm u}^\mu + \varepsilon {\bm T} {\bm\Delta}^{\mu\lambda} b_\lambda \,.
\end{align}
The resulting energy-momentum tensor and the current evaluated with the first-order distribution function (\ref{eq:ff-22}) are:
\begin{align}
\label{eq:T-1}
  &  T^{\mu\nu} = (T^{\mu\nu}_{0} + \varepsilon T^{\mu\nu}_{0,1}) + \varepsilon T^{\mu\nu}_1  = {\bf T}^{\mu\nu}[ T_{(0)} + \varepsilon T_{(1)} , \alpha_{(0)} + \varepsilon \alpha_{(1)}, u_{(0)} + \varepsilon u_{(1)}] + \varepsilon T^{\mu\nu}_1 \,,\\
\label{eq:J-1}
  &  J^\mu = (J^{\mu}_{0} + \varepsilon J^{\mu}_{0,1}) + \varepsilon J^{\mu}_1 = {\bf J}^\mu[T_{(0)} + \varepsilon T_{(1)}, \alpha_{(0)} + \varepsilon \alpha_{(1)}, u_{(0)} + \varepsilon u_{(1)}] + \varepsilon J^\mu_1 \,,
\end{align}
where the corrections $T^{\mu\nu}_1$, $J^\mu_1$ are due to the inhomogeneous solution $\Phi(x,p)$ in eq.~(\ref{eq:ff-22}).

\subsubsection*{First order: Inhomogeneous solution}
The hard part is to find the inhomogeneous solution $\Phi$ which satisfies
\begin{align}
\label{eq:BE-4}
  \frac{p^\mu \partial_\mu \f_p}{\f_p} = (1\pm \f_p) \left( p^\mu p^\nu \partial_\mu {\bm\beta}_\nu + p^\mu \partial_\mu {\bm\alpha} \right) = {\cal L}[\Phi]\,.
\end{align}
In general, for any timelike $\beta^\mu(x)$ and $\alpha(x)$ we have the identity
\begin{align}
\label{eq:ppdb}
  p^\mu p^\nu \partial_\mu {\beta}_\nu  + p^\mu \partial_\mu {\alpha}
  & = \frac{p^\mu p^\nu \sigma_{\mu\nu}[u] }{2T} + p^\mu \left( \partial_\mu^\perp \alpha - \frac{p{\cdot}u}{T}  (\partial_\mu^\perp T/T + \dot u_\mu)  \right) \nonumber\\
  & + \left( \frac{\pt^2}{dT} \partial{\cdot}u + \frac{(p{\cdot}u)^2}{T} \frac{\dot T}{T} - (p{\cdot}u) \dot \alpha \right) \,,
\end{align}
where $\sigma^{\mu\nu}[u] = (\Delta^{\mu\alpha}\Delta^{\nu\beta} + \Delta^{\nu\alpha}\Delta^{\mu\beta} -\frac{2}{d} \Delta^{\mu\nu}\Delta^{\alpha\beta})\partial_\alpha u_\beta$ is the shear tensor, $d$ is the number of space dimensions, $p^\mu_\perp\equiv \Delta^{\mu\lambda}p_\lambda$, and the hydrodynamic variables are $T = \frac{1}{\sqrt{-{\beta}^2}}$, $u^\mu = \frac{{\beta}^\mu}{\sqrt{-{\beta}^2}}$.

Based on eqs.~(\ref{eq:BE-4}) and (\ref{eq:ppdb}), it is difficult to guess how the solution $\Phi$ depends on the derivatives of the hydrodynamic variables, e.g.\ the relative contributions of $\dot {\bm T}$ and $\partial{\cdot}{\bm u}$ to $\Phi$. The standard approach is to use the constraints (\ref{eq:JT-cons-1}) to eliminate the zeroth-order $\dot {\bm u}_\mu$, $\dot {\bm T}$, and $\dot{\bm\alpha}$ in terms of the zeroth-order $\partial{\cdot}{\bm u}$, $\partial_\mu^\perp {\bm\alpha}$, $\partial_\mu^\perp {\bm T}$. Then the linearized Boltzmann equation (\ref{eq:BE-4}) becomes
\begin{align}
\label{eq:BE-5}
   (1\pm \f_p) \left[ F_\sigma p^\mu p^\nu \sigma_{\mu\nu}[{\bm u}] + F_u \partial{\cdot}{\bm u} + F_\alpha p^\mu \partial_\mu^\perp {\bm\alpha} \right] = {\cal L}[\Phi] \,,
\end{align}
where the functions $F_\sigma$, $F_u$, $F_\alpha$ depend on ${\bm T}$, ${\bm\alpha}$, $(p{\cdot}{\bm u})$, and are fixed by the ideal-gas equation of state. In particular, $F_\sigma = \frac{1}{2{\bm T}}$, $F_\alpha = 1+\frac{{\bm n}}{{\bm\epsilon}+{\bm p}}(p{\cdot}{\bm u})$. For massless particles, $F_u$ would vanish (at order $\varepsilon$), as a consequence of scale-invariant thermodynamics, $ {\bm p}({\bm T}, {\bm\alpha}) = {\bm T}^{d+1} g({\bm \alpha})$. Note that $\partial_\mu^\perp {\bm T}$ does not appear in the left-hand side of eq.~(\ref{eq:BE-5}), once $\dot {\bm u}_\mu$ has been eliminated. Now from eq.~(\ref{eq:BE-5}), the unknown $\Phi$ can be parametrized as 
\begin{align}
\label{eq:Phi-1}
  (1\pm \f_p) \Phi = K_\eta p^\mu p^\nu \sigma_{\mu\nu}[{\bm u}] + K_\zeta \partial{\cdot}{\bm u} + K_\alpha p^\mu \partial_\mu^\perp {\bm\alpha}\,,
\end{align}
where the coefficients $K_\eta$, $K_\zeta$, $K_\alpha$ in general depend on ${\bm T}$, ${\bm\alpha}$, and $p{\cdot}{\bm u}$, and can in principle be found by solving the linearized Boltzmann equation (\ref{eq:BE-5}). 
Let us write the first-order distribution function in terms of the first-order hydrodynamic variables $\beta_\mu = \beta_\mu^{(0)} + \varepsilon \beta_\mu^{(1)}$, $\alpha = \alpha^{(0)} + \varepsilon \alpha^{(1)}$,
\begin{align}
\label{eq:f-1a}
  f_p = \f_p (\beta, \alpha) + \varepsilon \, \f_p(\beta, \alpha) \left[ K_\eta p^\mu p^\nu \sigma_{\mu\nu}[u] + K_\zeta \partial{\cdot}u + K_\alpha p^\mu \partial_\mu^\perp \alpha \right] \,.
\end{align}
The first term has both $O(1)$ and $O(\varepsilon)$ contributions. In the second term, the $O(\varepsilon)$ contributions in $u$, $T$, and $\alpha$ give $O(\varepsilon^2)$ contributions to $f_p$ which can be neglected at first order. We can now use the distribution function (\ref{eq:f-1a}) to evaluate the corrections $T^{\mu\nu}_1$, $J^\mu_1$ in (\ref{eq:T-1}), (\ref{eq:J-1}).

\subsubsection*{First order: Constitutive relations}
Beyond leading (perfect-fluid) order, the energy-momentum tensor and the current will no longer have the simple form (\ref{eq:JT-0}). For any normalized timelike vector $u^\mu$, the energy-momentum tensor and the current may be decomposed as~\cite{Kovtun:2012rj}
\begin{align}
\label{eq:Td}
  & T^{\mu\nu} = {\E} u^\mu u^\nu + {\P} \Delta^{\mu\nu} + {\Q}^\mu u^\nu + {\Q}^\nu u^\mu + {\T}^{\mu\nu}\,,\\
\label{eq:Jd}
  & J^\mu = {\N} u^\mu + {\J}^\mu\,,
\end{align}
where ${\Q}{\cdot}u = {\J}{\cdot}u = {\T}{\cdot}u=0$, and ${\cal T}^{\mu\nu}$ is symmetric and traceless. These decompositions define $\E$, $\P$, $\Q$, $\T$, $\N$ and $\J$, for a given $u^\mu$. 
At first order in the $\varepsilon$-expansion,  $u^\mu = {\beta}^\mu/{\sqrt{-{\beta}^2}}$, where $\beta^\mu = \beta^\mu_{(0)} + \varepsilon \beta^\mu_{(1)}$, as in Eq.~(\ref{eq:u-01}). Similarly, at first order $T=T_{(0)} + \varepsilon T_{(1)}$, and $\alpha = \alpha_{(0)} + \varepsilon \alpha_{(1)}$. The first-order corrections to $\beta^\mu$ and $\alpha$ are arbitrary, and one can always redefine $\beta^\mu_{(1)} \to \beta^\mu_{(1)} + b'^\mu$, $\alpha_{(1)} \to \alpha_{(1)} + a'$.
At zeroth order in the expansion, $\E = {\bm\epsilon} + O(\varepsilon)$, $\P = {\bm p}+ O(\varepsilon)$, $\N = {\bm n} + O(\varepsilon)$, while $\Q^\mu$, $\T^{\mu\nu}$ and $\J^\mu$ are $O(\varepsilon)$.
Substituting the distribution function (\ref{eq:f-1a}) into the general expressions (\ref{eq:JJ}), (\ref{eq:TT}), we find the following coefficients of the decomposition (\ref{eq:Td}), (\ref{eq:Jd}) in terms of first-order $u^\mu$, $T$, and $\alpha$:
\begin{subequations}
\label{eq:cr-1}
\begin{align}
  & \E =  \epsilon + \varepsilon(\partial\epsilon/\partial T)  T^2 u{\cdot}b' + \varepsilon (\partial \epsilon/\partial \alpha) a' + \varepsilon \langle (p{\cdot}u)^2 K_\zeta \rangle \partial{\cdot}u + O(\varepsilon^2) \,,\\
  & \P =  p + \varepsilon(\partial p/\partial T)  T^2 u{\cdot}b' + \varepsilon (\partial p/\partial \alpha) a'  + \varepsilon  \langle \coeff{1}{d} \pt^2 K_\zeta \rangle \partial{\cdot}u + O(\varepsilon^2)\,,\\
  & \Q^\mu = \varepsilon(\epsilon{+}p) T \Delta^{\mu\lambda} b'_\lambda - \varepsilon \langle \coeff{1}{d} \pt^2 (p{\cdot}u) K_\alpha \rangle  \partial^\mu_\perp \alpha + O(\varepsilon^2) \,,\\
  & \T^{\mu\nu} = \varepsilon \coeff{2}{d(d+2)} \langle (\pt^2)^2 K_\eta \rangle \sigma^{\mu\nu} + O(\varepsilon^2)\,,\\
  & \N = n + \varepsilon (\partial n/\partial T)  T^2 u{\cdot}b' + \varepsilon (\partial n/\partial \alpha) a' - \varepsilon \langle p{\cdot}u\, K_\zeta \rangle \partial{\cdot}u + O(\varepsilon^2)\,,\\
  & \J^\mu = \varepsilon n T \Delta^{\mu\lambda} b'_\lambda + \varepsilon  \langle \coeff{1}{d} \pt^2 K_\alpha \rangle  \partial^\mu_\perp \alpha + O(\varepsilon^2) \,.
\end{align}
\end{subequations}
Here $\epsilon$, $p$, and $n$ are functions of ($\varepsilon$-corrected) $T$ and $\alpha$. The angular brackets stand for $\langle\cdots\rangle = \int_p \f_p \cdots$. These are the constitutive relations for a viscous relativistic fluid at first order in the derivative expansion. The energy-momentum tensor and the current given by these constitutive relations (in terms of $O(\varepsilon)$-corrected hydrodynamic variables) must obey the standard conservation equations (\ref{eq:JT-cons}), which are true for any distribution function.

\subsubsection*{First order: Hydrodynamic ``frames''}
One might be tempted to ignore the ``integration constants'' $b_\mu'(x)$ and $a'(x)$ altogether. However, they have a simple physical meaning: the hydrodynamic variables $T$, $u^\lambda$, and $\alpha$ that appear in the $O(\varepsilon)$ (i.e.\ Navier-Stokes) hydrodynamic equations can differ from the hydrodynamic variables that appear in the distribution function (\ref{eq:ff}) by derivative corrections, reflecting the ambiguity in what one chooses to mean by ``fluid velocity'', ``fluid temperature'' and ``fluid chemical potential'' beyond the perfect-fluid approximation. The most general parametrization of such arbitrary one-derivative corrections is%
\begin{subequations}
\label{eq:bb-aa-1}
\begin{align}
\label{eq:bb-1}
  & b'_\mu  = (b_1 \dot T/T + b_2 \partial{\cdot}u + b_3 \dot\alpha) u_\mu + b_4 \dot u_\mu + b_5 \partial_\mu^\perp T /T + b_6 \partial_\mu^\perp \alpha\,,\\
\label{eq:aa-1}
  & a' = a_1 \dot T/T + a_2 \partial{\cdot}u + a_3 \dot\alpha\,,
\end{align}
\end{subequations}
with arbitrary coefficients $b_n(T,\alpha)$ and $a_n(T,\alpha)$. 
In relativistic hydrodynamics, one's choice of a particular form of these derivative corrections is often called a choice of ``frame''. 

The parametrization (\ref{eq:bb-aa-1}) contains the most general one-derivative corrections with arbitrary coefficients $b_n$ and $a_n$. One could further demand that the redefinitions of $T$, $\alpha$ and $u^\lambda$ (provided by $b'_\mu$ and $a'$) are such that they vanish in equilibrium, even when the fluid is subject to a static external gravitational field. In equilibrium, one can choose the fluid velocity as the normalized timelike Killing vector. In zero-derivative hydrodynamics (perfect fluids) this is manifested by eq.~(\ref{eq:ba-eq}), however such a choice of the fluid velocity {\em in equilibruim} of course extends beyond zero-derivative hydrodynamics, and has non-trivial consequences~\cite{Jensen:2012jh}. The Killing equation (\ref{eq:ba-eq}) for $\beta_\mu$ implies $\dot u_\mu + \partial_\mu^\perp T/T = 0$, even though $\dot u_\mu$ and $\partial_\mu^\perp T$ may separately be non-zero in external gravitational field. Thus demanding that (\ref{eq:bb-1}), (\ref{eq:aa-1}) vanish in equilibrium, we have $b_4=b_5$. Such a choice was called a ``thermodynamic frame'' in ref.~\cite{Jensen:2012jh}. The choice amounts to demanding that the hydrostatic limit of the constitutive relations (\ref{eq:cr-1}) follows by varying the equilibrium grand canonical free energy with respect to the external metric (for $T^{\mu\nu}$), or with respect to the external gauge field (for $J^\mu$). 

The popular frame adopted by Landau and Lifshitz~\cite{LL6} is obtained in the following way. 
One chooses $b_1=b_3 = b_4 = b_5 = a_1 = a_3 = 0$, leaving one with
\begin{align}
  b'_\mu =  b_2 (\partial{\cdot}u) u_\mu +  b_6 \partial_\mu^\perp \alpha \,,\ \ \ \ 
  a' =  a_2 (\partial{\cdot}u) \,.
\end{align}
The arbitrary coefficients $ b_2$ and $ a_2$ are fixed by demanding that ${\cal E} = \epsilon + O(\varepsilon^2)$, ${\cal N} = n + O(\varepsilon^2)$. Following the constitutive relations (\ref{eq:cr-1}), this determines $ b_2$ and $ a_2$ in terms of $\langle (p{\cdot}u)^2 K_\zeta \rangle$ and $\langle (p{\cdot}u) K_\zeta \rangle$. After that, the non-equilibrium pressure takes the form ${\cal P} = p - \varepsilon \zeta (\partial{\cdot}u) + O(\varepsilon^2)$, where $\zeta$ is the bulk viscosity,
\begin{align}
\label{eq:zeta-0}
  \zeta = \left( \left(\frac{\partial p}{\partial\epsilon}\right)_{\!n} -\frac1d \right) \langle (p{\cdot}u)^2 K_\zeta \rangle - \left(\frac{\partial p}{\partial n}\right)_{\!\epsilon} \langle (p{\cdot}u) K_\zeta \rangle + \frac{m^2}{d} \langle K_\zeta \rangle\,,
\end{align}
and we have used the on-shell relation $m^2 = (p{\cdot}u)^2 - \pt^2$. For massless particles, we have $m^2=0$, $\epsilon = d\, p$, and the above expression gives $\zeta=0$. Finally, the coefficient $ b_6$ is fixed by demanding ${\cal Q}^\mu = O(\varepsilon^2)$. Following the constitutive relations (\ref{eq:cr-1}), this determines $ b_6$ in terms of $\langle  \pt^2 (p{\cdot}u) K_\alpha \rangle$. The particle number flux takes the form ${\cal J}^\mu = - \varepsilon \sigma T \partial^\mu_\perp \alpha + O(\varepsilon^2)$, where $\sigma$ is the particle number conductivity (which would become electrical conductivity if the particles were to carry electric charge),
\begin{align}
\label{eq:sigma-0}
  \sigma = -\frac{1}{d\, T} \langle \pt^2 K_\alpha \rangle - \frac{1}{d\, T} \frac{n}{\epsilon{+}p} \langle \pt^2 (p{\cdot}u) K_\alpha \rangle\,.
\end{align}

The frame of Eckart \cite{PhysRev.58.919} is obtained in a similar manner. One chooses $b_4=b_5$ (consistent with the thermodynamic frame), and sets $b_1 = b_3 = b_6 = a_1 = a_3 = 0$, so that
\begin{align}
  b'_\mu =  b_2 (\partial{\cdot}u) u_\mu +  b_4 \left( \dot u_\mu + \frac{\partial_\mu^\perp T}{T} \right) \,,\ \ \ \ 
  a' = a_2 (\partial{\cdot}u) \,.
\end{align}
The arbitrary coefficients $ b_2$ and $ a_2$ are fixed by demanding that ${\cal E} = \epsilon + O(\varepsilon^2)$, ${\cal N} = n + O(\varepsilon^2)$, while $ b_4$ is fixed by demanding ${\cal J}^\mu = O(\varepsilon^2)$. The bulk viscosity $\zeta$ again arises as the non-equilibrium correction to pressure, while the conductivity $\sigma$ arises as the non-equilibrium contribution to the energy flux ${\cal Q}^\mu$.

The transport coefficients $\zeta$ and $\sigma$ are physical observables, and do not depend on how one chooses to fix the arbitrary coefficients in eq.~(\ref{eq:bb-aa-1}). For example, one could choose a frame where the bulk viscosity arises as a non-equilibrium correction to the energy density, while the pressure stays uncorrected to first order, ${\cal P} = p + O(\varepsilon^2)$. The actual values of $\zeta$ and $\sigma$ are of course unchanged by where they appear in the constitutive relations~\cite{Kovtun:2019hdm}.

In the above examples of Landau-Lifshitz and Eckart frames, the arbitrary coefficients $b_n(T,\alpha)$ and $a_n(T,\alpha)$ in eq.~(\ref{eq:bb-aa-1}) were fixed by a choice of aesthetics. For example, in the Landau-Lifshitz frame the fluid velocity $u^\mu$ appears as an eigenvector of the energy-momentum tensor, while in the Eckart frame the equations resemble the historical formulation of the non-relativistic equations of compressible dissipative hydrodynamics. The idea behind BDNK hydrodynamics is: rather than being guided by aesthetics, the arbitrary coefficients $b_n$ and $a_n$ need to be chosen in a way that makes the resulting hydrodynamical equations mathematically well-posed.
It is a non-trivial statement that it {\em is} in fact possible to choose the coefficients $a_n$, $b_n$ such that the hydrodynamic equations are hyperbolic and causal. We illustrate this in appendices~\ref{ap::concrete} and \ref{sec:realspace}.

\subsubsection*{Second order}
Going to order $O(\varepsilon^2)$ the Boltzmann equation becomes
\begin{align}
\label{eq:BE-22}
   \varepsilon \, p^\mu \partial_\mu  \left( \f + \varepsilon f^{(1)}\right)= \C[\f + \varepsilon f^{(1)} + \varepsilon^2 f^{(2)}]\,.
\end{align}
Recall that the linearized collision operator is defined as $\C[\f + \delta\!f] = \f {\cal L}[\phi] + O(\delta\! f^2)$, where $\phi = \delta \! f/\f(1\pm\f)$. Thus to order~$O(\varepsilon^2)$ we have 
\begin{align}
  \C[\f + \varepsilon f^{(1)} + \varepsilon^2 f^{(2)}] = \f {\cal L}[\varepsilon\phi^{(1)} + \varepsilon^2 \phi^{(2)}] + \varepsilon^2 C^{(2)}[\phi^{(1)}],
\end{align}
where $\phi^{(n)} \equiv f^{(n)}/\f(1\pm\f)$, and $C^{(2)}$ is quadratic in $\phi^{(1)}$, but does not contain $\phi^{(2)}$.
Without specifying the explicit form of $C^{(2)}$, it follows that for arbitrary $a(x)$, $b_\mu(x)$ we have
\begin{align}
\label{eq:CC-2}
  \int_p \left( a(x) + b_\mu(x) p^\mu \right) C^{(2)}[\phi^{(1)}_p] = 0,
\end{align}
as a consequence of the microscopic conservation laws embodied by eqs.~(\ref{eq:C-cons}), (\ref{eq:L-zeromodes}), and (\ref{eq:L-symmetry}). The Boltzmann equation at order $O(\varepsilon^2)$ is 
\begin{align}
\label{eq:BE-6}
  p^\mu \partial_\mu f^{(1)} - C^{(2)}[\phi^{(1)}] = \f {\cal L}[\phi^{(2)}]\,,
\end{align}
where $f^{(1)}$ in the left-hand side is known from the $O(\varepsilon)$ calculation in eqs.~(\ref{eq:ff-22}), (\ref{eq:Phi-1}),
\begin{align}
\label{eq:f-1}
  f^{(1)}_p = \f_p(1\pm \f_p) \left( a^{(1)}+b^{(1)}_\mu p^\mu \right) + \f_p \left( K_\eta p^\mu p^\nu \sigma_{\mu\nu} + K_\zeta \partial{\cdot}u + K_\alpha p^\mu \partial_\mu^\perp \alpha \right)\,.
\end{align}
As before, the linear equation (\ref{eq:BE-6}) can only be solved for $\phi^{(2)}$ if the left-hand side is orthogonal to the zero-modes of the operator ${\cal L}$ in the right-hand side. The quadratic part $C^{(2)}$ drops out from the orthogonality condition thanks to eq.~(\ref{eq:CC-2}), and the constraint becomes $\int_p \left( a + b_\nu p^\nu \right) p^\mu \partial_\mu f^{(1)} = 0$, or equivalently 
\begin{align}
  \partial_\mu J^\mu_{(1)} = 0\,,\ \ \ \ \partial_\mu T^{\mu\nu}_{(1)} = 0\,.
\end{align}
Here $J^\mu_{(1)}$ and $T^{\mu\nu}_{(1)}$ are given by eqs.~(\ref{eq:JJ}), (\ref{eq:TT}), evaluated with $f^{(1)}_p$ in eq.~(\ref{eq:f-1}). Connecting these expressions to $T^{\mu\nu}_{(1)}$ in eq.~(\ref{eq:TT-1}), the first term in (\ref{eq:f-1}) gives $T^{\mu\nu}_{0,1}$, the second term in (\ref{eq:f-1}) gives $T^{\mu\nu}_1$, and similarly for the current $J^\mu_{(1)}$.
In other words, the equations of 1-st~order (Navier-Stokes) hydrodynamics arise as constraint equations at 2-nd~order in the expansion.
The same happens to all orders: the equations (\ref{eq:TJn-cons}) of $n^{\rm th}$-order hydrodynamics arise as constraint equations at $(n{+}1)^{\rm th}$-order in the expansion.

\section{Fluid/Gravity correspondence}\label{sec::fg-correspond}
\noindent
In the preceding section, we have outlined a procedure to derive causal hydrodynamics from kinetic theory. There is an analogous procedure to derive hydrodynamic equations from classical gravity in asymptotically anti-de Sitter spacetimes. This is done via the fluid/gravity correspondence \cite{Bhattacharyya:2008jc, Erdmenger:2008rm, Banerjee:2008th}, see \cite{Hubeny:2011hd} for a review.

\subsection*{Einstein-Maxwell equations and Hilbert expansion}
\noindent
Following the original fluid-gravity discussion, we focus on the simplest holographic model of a 3+1 dimensional quantum field theory with a conserved global U(1) symmetry: the Einstein-Maxwell theory in AdS$_{\rm 5}$,
\begin{equation}\label{eq:EM-action}
S =  \frac{1}{16 \pi G_N^{(5)}} \int d^5x \sqrt{-g} \biggl[ R + 12 - \frac{1}{4}F_{MN} F^{MN}\biggr],
\end{equation}
where latin indices $M,N$ are bulk indices; greek indices, raised and lowered by the Minkowski metric $\eta^{\mu\nu}$, will be used for the boundary directions. The AdS radius of curvature has been set to one, hence the cosmological constant is $\Lambda = -6$. 
The Einstein-Maxwell equations are
\begin{subequations}\label{eq:EM-eqs}
\begin{align}
  &R_{MN} - \frac{1}{2} R g_{MN} - 6 g_{MN} + \frac{1}{2} \biggl[ F_{MK} F^{K}_{\,\,N} + \frac{1}{4} g_{MN} F_{KL} F^{KL}\biggr] = 0,\\
  &\nabla_{M} F^{MN} = 0.
\end{align}
\end{subequations}
The solution of \eqref{eq:EM-eqs} that corresponds to the equilibrium state in the dual field theory at non-zero temperature and non-zero U(1) charge density is the electrically charged black brane,%
\begin{subequations}\label{eq:RN-sol-eq}
\begin{align}
&ds^2 = -2 u_\mu dx^\mu dr - r^2 f(r) u_\mu u_\nu dx^\mu dx^\nu + r^2 \Delta_{\mu\nu} dx^\mu dx^\nu, \quad f(r) = 1 - \frac{1}{b^4r^4} + \frac{Q^2}{r^6},\\
&A_M dx^M = \frac{\sqrt{3} Q}{2 r^2} u_\mu dx^\mu.
\end{align}
\end{subequations}
 The solution contains three constant parameters: a timelike covector~$u_\mu$ (normalized such that $u_\mu u^\mu = -1$), a charge $Q$, and a mass parameter $b$. As before, $\Delta_{\mu\nu} = \eta_{\mu\nu}+ u_\mu u_\nu$ is the spatial projector on the boundary. This metric is written in infalling Eddington-Finkelstein coordinates. The vector $u^\mu$ defines the rest frame of the fluid on the boundary. The parameters $b$ and $Q$ are (somewhat unilluminating) functions of the temperature $T$ and the U(1) chemical potential $\mu$ of the boundary fluid. The explicit expressions for $b(T,\mu)$ and $Q(T,\mu)$ can be obtained from refs.~\cite{Erdmenger:2008rm, Banerjee:2008th}, in particular $b(T,\mu{\to}0) = 1/\pi T$ and $Q(T,\mu{\to}0)=0$.

Drawing an analogy with kinetic theory, the equilibrium metric $\g$ and the equilibrium gauge field $\A$ of eq.~\eqref{eq:RN-sol-eq} are the holographic analogues of the equilibrium distribution function $\f$. If  the parameters $b$, $Q$, and $u_\mu$ are promoted to be functions of the boundary coordinates, i.e.\ $b(x), u_\mu(x), Q(x)$, then \eqref{eq:RN-sol-eq} is no longer a solution to \eqref{eq:EM-eqs}. However, in analogy with kinetic theory, we may construct approximate solutions through a Hilbert expansion of the form
\begin{subequations}\label{eq:EM-hilbert}
\begin{align}
&g(x) = \g(x) + \ce g^{(1)}(x) + \ce^2 g^{(2)}(x) + O(\ce^3),\\
&A(x) = \A(x) + \ce A^{(1)}(x) + \ce^2 A^{(2)}(x) + O(\ce^3).
\end{align}
\end{subequations}
Similarly, the parameters themselves get corrected order-by-order as well:
\begin{subequations}\label{eq:EM-buq-correct}
\begin{align}
&b(x) = b^{(0)}(x) + \ce b^{(1)}(x) + \ce^2 b^{(2)}(x) + O(\ce^3),\\
&u_\mu(x) = u_\mu^{(0)}(x) + \ce u_{\mu}^{(1)}(x) + \ce^2 u_\mu^{(2)}(x) + O(\ce^3),\\
&Q(x) = Q^{(0)}(x) + \ce Q^{(1)}(x) + \ce^2 Q^{(2)}(x) + O(\ce^3).
\end{align}
\end{subequations}
Inserting \eqref{eq:EM-hilbert} into the Einstein-Maxwell equations \eqref{eq:EM-eqs} and equating like-powers gives an analogue to the linearized Boltzmann equation:
\begin{equation}\label{eq:EE-linearized}
\H[g^{(n)}, A^{(n)}] = s_n[g^{(n-1)}, A^{(n-1)}, ...],
\end{equation}
where the operator $\H$, like the linearized collision operator $\L$, depends only of the equilibrium metric $\g$ and equilibrium gauge field $\A$, is the same at all orders in $\ce$, and (crucially) has zero modes. In the same way that $\L$ involves integrals of $p$, the operator $\H$ involves derivatives with respect to $r$ (compare with the interpretation of the $r$-direction as the energy scale in the dual field theory). The source term depends only on the lower-order corrections to the metric and the gauge field. The explicit expressions for $\H$ and $s_0$, $s_1$, $s_2$ may be found in~\cite{Bhattacharyya:2008jc, Erdmenger:2008rm, Banerjee:2008th}. The constraint equations in the bulk give rise to 
\begin{equation}\label{eq:EE-constraint}
\partial_\mu T^{\mu\nu}_{(n-1)} = 0, \qquad \partial_\mu J^\mu_{(n-1)} = 0,
\end{equation}
where $T^{\mu\nu}_{(n-1)}$ and $J^{\mu}_{(n-1)}$ are the $(n-1)^{\rm th}$-order correction to the boundary stress-energy tensor and the U(1) charge current, respectively. Again, this is the exact same constraint that one finds in kinetic theory: the perfect-fluid equations come about as a constraint at first order, the Navier-Stokes equations arise as a constraint at second order, etc.

\subsection*{Zero modes of $\H$}
\noindent
The operator $\H$ is a linear differential operator in $r$ whose coefficients depend on the zeroth-order functions $b^{(0)}(x)$, $u_\mu^{(0)}(x)$, and $Q^{(0)}(x)$. Hence, the zero modes of $\H$ must be specific functions of $r$ and $b^{(0)}(x)$, $u_\mu^{(0)}(x)$, $Q^{(0)}(x)$ which are multiplied by arbitrary functions of the boundary coordinates (integration constants with respect to $r$).

Suppose we want to solve the Einstein-Maxwell equations to first order in $\ce$. After enforcing boundary conditions and removing gauge redundancies, the solution can be written as%
\begin{subequations}\label{eq:EM-sol}
\begin{align}
&ds^2 = \g_{MN} dx^M dx^N \nonumber \\
&\quad\,\,\,+ \ce \biggl( G^{(1)}_{MN} + {\frak b}(x) f^{(b)}_{MN} + {\frak q}(x) f^{(Q)}_{MN}+  {\frak u}_{\lambda}(x) \left[f^{(u)}\right]^{\lambda}_{MN} \biggr)dx^M dx^N + O(\ce^2),\\
&A_M dx^M = \A_M dx^M + \ce \left( A^{(1)}_{M} dx^M+ {\frak q}(x) f^{(Q)}_{M} + {\frak u}_{\lambda}(x)  \left[f^{(u)}\right]^{\lambda}_{M}\right)dx^M + O(\ce^2)\,.
\end{align}
\end{subequations}
Here $G_{MN}^{(1)}$, $A_M^{(1)}$  are particular solutions found by inverting the operator $\H$ (analogous to $\Phi_p$ in the previous section), and depend on the source $s_1$ in the right-hand side of \eqref{eq:EE-linearized}. The functions ${\frak b}(x)$, ${\frak q}(x)$, and ${\frak u}_\lambda(x)$ are ``integration constants" with respect to $r$, and $u^\lambda {\frak u}_\lambda = 0$. Finally, $f^{(b)}_{MN}$, $f^{(Q)}_{MN}$, $\left[f^{(u)}\right]^{\lambda}_{MN}$, $f_M^{(Q)}$, and $\left[f^{(u)}\right]^\lambda_M$ are the accompanying bases for the zero modes of $\H$. Their explicit form is
\begin{subequations}\label{eq:holo-zeromodes}
\begin{align}
f_{MN}^{(b)} dx^M dx^N &=-\frac{4}{b_{(0)}^5 r^2} u_\mu^{(0)} u_\nu^{(0)} dx^\mu dx^\nu,\\
f_{MN}^{(Q)} dx^M dx^N &= - \frac{2 Q^{(0)}}{r^4} u_\mu^{(0)} u_\nu^{(0)} dx^\mu dx^\nu,\\
\biggl[f^{(u)}\biggr]^{\lambda}_{MN} dx^M dx^N &= r^2 (1-f^{(0)}(r)) \eta^{\lambda \alpha} \left(\Delta_{\alpha\mu}^{(0)} u_\nu^{(0)} + \Delta_{\alpha\nu}^{(0)} u_\mu^{(0)}\right) dx^\mu dx^\nu,\\
f_M^{(Q)} dx^M &= \frac{\sqrt{3}}{2 r^2} u_\mu^{(0)} dx^\mu,\\
\biggl[f^{(u)}\biggr]^{\lambda}_{M} dx^M &= \frac{\sqrt{3}Q^{(0)}}{2 r^2} \eta^{\lambda\alpha} \Delta_{\alpha\mu}^{(0)} dx^\mu,
\end{align}
\end{subequations}
where $f^{(0)}(r) = 1 - 1/(b_{(0)}^4 r^4) + Q_{(0)}^2/r^6$. The normalization in \eqref{eq:holo-zeromodes} has been chosen so that the ``integration constants'' ${\frak b}$, ${\frak q}$, and ${\frak u}_{\lambda}$ represent corrections to the quantities $b$, $Q$, and $u_\mu$, respectively. We can see that this is the case by looking at the equilibrium metric $\g$ and equilibrium gauge field $\A$, and then expanding these parameters as in \eqref{eq:EM-buq-correct}. Note that $u_{\lambda}^{(1)} u^{\lambda}_{(0)} = O(\ce^2)$, and so to first order in $\ce$, $u^{\lambda}_{(1)} \Delta_{\lambda\mu}^{(0)} = u_{\mu}^{(1)}$. Expanding, we find
\begin{align}
  \g_{MN} dx^M dx^N = &-2 u_\mu^{(0)} dx^\mu dr - r^2 \left(1 - \frac{1}{b_{(0)}^4 r^4} + \frac{Q_{(0)}^2}{r^6}\right) u^{(0)}_\mu u^{(0)}_\nu dx^\mu dx^\nu + r^2 \Delta^{(0)}_{\mu\nu} dx^\mu dx^\nu\nonumber\\
& + \ce \biggl[ -2 u^{(1)}_\mu dx^\mu dr - \left( \frac{4 b^{(1)}}{r^2 b_{(0)}^5} + \frac{2 Q^{(0)} Q^{(1)}}{r^4}\right) u_\mu^{(0)} u_\nu^{(0)} dx^\mu dx^\nu \nonumber\\
&+r^2 (1-f^{(0)}(r)) u_{\alpha}^{(1)} \eta^{\alpha\lambda}\left( \Delta_{\lambda\mu}^{(0)} u_\nu^{(0)} + u_\mu^{(0)} \Delta_{\lambda\nu}^{(0)}\right)dx^\mu dx^\nu\biggr] + O(\ce^2),
\end{align}
\begin{align}
  \A_{M} dx^M = \frac{\sqrt{3} Q^{(0)}}{2 r^2} u_\mu^{(0)} dx^\mu + \ce \biggl[ \frac{\sqrt{3} Q^{(1)}}{2 r^2} u_\mu^{(0)} dx^\mu + \frac{\sqrt{3} Q^{(0)}}{2 r^2} u_{\alpha}^{(1)} \eta^{\alpha\lambda} \Delta_{\lambda\mu}^{(0)} dx^\mu\biggr] + O(\ce^2).
\end{align}
By direct comparison, one can see that $b^{(1)} = {\frak b}(x)$, $Q^{(1)} = {\frak q}(x)$, and $u_\mu^{(1)} = {\frak u}_{\mu}(x)$. As the hydrodynamic variables $\beta_\mu=u^\mu/T$ and $\alpha=\mu/T$ are functions of $b$, $Q$, and $u_\mu$, the ``integration constants'' ${\frak b}$, ${\frak q}$, and ${\frak u}^\lambda$ will set the hydrodynamic ``frame''. The corrections to the ``conventional'' hydrodynamic variables $\alpha=\mu/T$ and $\beta_\mu = u_\mu/T$ are given by
\begin{subequations}
\begin{align}
\alpha^{(1)} &= \frac{\partial \alpha^{(0)}}{\partial b^{(0)}} {\frak b} + \frac{\partial\alpha^{(0)}}{\partial Q^{(0)}} \,{\frak q},\\
\beta_\mu^{(1)} &= - \frac{1}{T_{(0)}^2} \left( \frac{\partial T^{(0)}}{\partial b^{(0)}} \,{\frak b} + \frac{\partial T^{(0)}}{\partial Q^{(0)}} {\frak q} \right) u_\mu^{(0)} + \frac{1}{T_{(0)}} {\frak u}^{\lambda} \Delta_{\lambda \mu}^{(0)}\,.
\end{align}
\end{subequations}
The partial derivatives can be evaluated by inverting the known equilibrium functions $b(T,\mu)$ and $Q(T,\mu)$ to find $T(b,Q)$ and $\mu(b,Q)$. Thus fixing ${\frak b}$, ${\frak q}$, and ${\frak u}_\mu$ is equivalent to fixing the definitions of the hydrodynamic variables $\beta_\mu$ and $\alpha$ at $O(\ce)$. The original fluid-gravity references \cite{Bhattacharyya:2008jc, Erdmenger:2008rm, Banerjee:2008th} adopted the Landau-Lifshitz convention, however tuning ${\frak b}$, ${\frak q}$, and ${\frak u}_\mu$ may be used to generate other conventions. In particular, hydrodynamic field redefinitions can be used to arrive at stable and causal first-order hydrodynamics as described in~\cite{Kovtun:2019hdm}.

\section{Conclusions} \label{sec::conc}
\noindent 
Physically, hydrodynamics is a theory of local densities of conserved quantities (energy, momentum, etc) which can not disappear through microscopic interactions, but rather spread out through the corresponding fluxes. On the other hand, when derived from a more fundamental microscopic description such as the kinetic theory or holography, classical hydrodynamics may be viewed as a theory of zero modes. In kinetic theory, the zero modes are those of the linearized collision operator~${\cal L}$. In the fluid-gravity correspondence, the zero modes are those of the operator $\H$. While the bulk fields in the fluid-gravity correspondence are the analogues of the distribution function, the operator $\H$ is the analogue of the linearized collision operator. 
Indeed, as was emphasized in ref.~\cite{Hubeny:2011hd}, the equations of bulk dynamics may be considered as a strong-coupling analogue of the Boltzmann equation.

The freedom of choosing the zero modes at each order of the derivative expansion translates to the freedom of field redefinitions of the hydrodynamic variables. While in kinetic theory the zero modes are naturally associated with the shifts of $\beta^\mu = u^\mu/T$ and $\alpha=\mu/T$ which parametrize the equilibrium distribution function, the zero modes in the fluid-gravity correspondence are naturally associated with the shifts of $b(T,\alpha)$, $u^\mu$, and $Q(T,\alpha)$ which parametrize the equilibrium bulk metric and the gauge field. Still, hydrodynamic field redefinitions work in exactly the same way in both setups: neither the Hilbert expansion in kinetic theory nor the analogous expansion in fluid-gravity come with a preferred ``frame''. In both kinetic theory and in fluid-gravity one may obtain causal hydrodynamic equations through a judicious choice of zero modes at one-derivative order. We plan to return to further exploring the connections between the Botlzmann equation and the fluid-gravity duality in the future.

\acknowledgments
This work was supported in part by the NSERC of Canada.

\appendix

\section{Examples of causal frames}
\label{ap::concrete}
\noindent
Following ref.~\cite{Kovtun:2019hdm}, we will denote the one-derivative terms in the constitutive relations as
\begin{subequations}
\label{eq:cr-2}
\begin{align}
&\E = \epsilon + \ce_1 \dot{T}/T + \ce_2 \partial{\cdot}u + \ce_3 \dot{\alpha} + O(\partial^2),\\
&\P = p + \pi_1 \dot{T}/T + \pi_2 \partial{\cdot}u + \pi_3 \dot{\alpha} + O(\partial^2),\\
&\Q^\mu = \theta_1 \left( \dot{u}^\mu + \partial^\mu_{\perp} T/T\right) + \theta_3 \partial^\mu_\perp \alpha + O(\partial^2),\\
&\T^{\mu\nu} = - \eta \sigma^{\mu\nu} + O(\partial^2),\\
&\N = n + \nu_1 \dot{T}/T + \nu_2 \partial{\cdot}u + \nu_3 \dot{\alpha} + O(\partial^2),\\
&\J^\mu = \gamma_1 \left( \dot{u}^\mu + \partial^\mu_{\perp} T/T\right) + \gamma_3 \partial^\mu_\perp \alpha + O(\partial^2).
\end{align}
\end{subequations}
The combinations of the transport parameters that are invariant under the redefinitions of $T$, $u^\mu$, and $\alpha$ by one-derivative corrections are \cite{Kovtun:2019hdm}
\begin{align}
\label{eq:fi-def}
  & f_i = \pi_i - \varepsilon_i (\partial p/\partial\epsilon)_n - \nu_i (\partial p/\partial n)_\epsilon\,,\\
\label{eq:li-def}
  & \ell_i = \gamma_i - n\theta_i /(\epsilon{+}p)\,.
\end{align}
The physical transport coefficients (bulk viscosity and charge conductivity) are
\begin{align}
  & \zeta = -f_2 + \left( \frac{\partial p}{\partial\epsilon} \right)_{\!n} f_1 + \frac{1}{T} \left( \frac{\partial p}{\partial n}\right)_{\!\epsilon} f_3\,,\\
  & \sigma = -\frac{1}{T} \ell_3 + \frac{n}{\epsilon+p} \ell_1 \,.
\end{align}
Suppose that we have found the distribution function at $O(\varepsilon)$ by eliminating $\dot{u}^\mu$, $\dot{T}$ and $\dot{\alpha}$, as in eq.~(\ref{eq:f-1a}). This choice of eliminating the time derivatives gives rise to the following constraints on the transport parameters:
\begin{subequations}
\label{eq:HE-c}
\begin{align}
&\pi_1 = \left(\pder{p}{\epsilon}\right)_n \ce_1 + \left( \pder{p}{n} \right)_\epsilon \nu_1,\\
&\pi_2 = \left(\pder{p}{\epsilon}\right)_n \ce_2 + \left( \pder{p}{n} \right)_\epsilon \nu_2 - \zeta,\\
&\pi_3 = \left(\pder{p}{\epsilon}\right)_n \ce_3 + \left( \pder{p}{n} \right)_\epsilon \nu_3,\\
&\gamma_1 = \frac{n}{\epsilon + p} \theta_1,\\
&\gamma_3 = \frac{n}{\epsilon + p} \theta_3 - T \sigma.
\end{align}
\end{subequations}
We also take $\theta_1=\theta_2$, $\gamma_1=\gamma_2$, as required in a thermodynamic frame.
In other words, eliminating the time derivatives makes the frame invariants $f_1$, $f_3$, $\ell_1$, $\ell_2$ vanish, while $f_2=-\zeta$, and $\ell_3 = -T\sigma$. The non-zero values of the transport parameters $\varepsilon_i$, $\pi_i$ etc in (\ref{eq:cr-2}) are due to the zero modes of the linearized collision operator, with the exception of the physical transport coefficients $\eta$, $\zeta$, and $\sigma$ which are of course insensitive to the zero modes.

Ignoring the kinetic theory motivation, one can simply view (\ref{eq:HE-c}) as a particular set of constraints which one may choose to impose on the one-derivative transport parameters. We will now show that these constraints are consistent with causality, in other words that one may choose transport parameters $\varepsilon_i$, $\pi_i$, etc such that the constraints (\ref{eq:HE-c}) are satisfied, and the hydrodynamic equations with the constitutive relations (\ref{eq:cr-2}) are causal.

We start with one-derivative hydrodynamics of conformal fluids. In $d$ space dimensions, $p{=}\epsilon/d$, hence $\left(\partial p/\partial n\right)_\epsilon = 0$, and $\nu_n$ drop out of the constraints \eqref{eq:HE-c}. Further, conformal symmetry dictates $\zeta=0$, $\pi_n = \varepsilon_n/d$, and the first three constraints in \eqref{eq:HE-c} are satisfied identically. Additionally, conformal symmetry implies $\ce_2 = \ce_1/d$, $\nu_2 = \nu_1/d$~\cite{Kovtun:2019hdm}. A class of causal and stable frames which satisfy (\ref{eq:HE-c}) was given in ref.~\cite{Hoult:2020eho}: choosing
\begin{align}
  & \varepsilon_3 = \theta_3 = \nu_1 = 0\,,\\
  & \nu_3 > \sigma T\,,\\
  & \varepsilon_2 > (2d-2)\eta\,,\\
  & 1 - \frac{2d}{d{-}1} \frac{\eta}{\theta_1} - \frac{2}{d{-}1} \frac{\eta}{\varepsilon_2} > 0\,,
\end{align}
will give rise to causal first-order hydrodynamics. In $d=3$ space dimensions, it suffices to choose $\nu_3>\sigma T$, $\varepsilon_2 >4\eta$, $\theta_1>4\eta$ in order to satisfy the above inequalities.

Now consider non-conformal fluids in $d=3$ space dimensions. As an example, suppose we narrow down the class of frames by demanding that
\begin{equation}
  \nu_1 = \left(\frac{n}{\epsilon + p}\right) \ce_1,\ \ \ \ 
  \nu_2 = \left(\frac{n}{\epsilon + p}\right) \ce_2, \ \ \ \ 
  \nu_3 = \delta + \left(\frac{n}{\epsilon + p}\right) \ce_3 ,
\end{equation}
where $\delta$ is to be constrained momentarily. The short-wavelength modes propagate with a linear dispersion relation $\omega({\bf k}) = \pm c_s |{\bf k}|$, where the speed $c_s$ is determined by
\begin{align}
  \left( c_s^2 \theta_1 - \eta \right)^2 \left(c_s^2 \delta - T\sigma \right) \left( c_s^4 - c_s^2 \left( v_s^2 {+} \frac{\varepsilon_2}{\varepsilon_1} {+} \frac{\gamma_s}{\theta_1} \right)  + \frac{v_s^2 \varepsilon_2 - \gamma_s}{\varepsilon_1} \right) = 0 \,,
\end{align}
where $v_s^2 = (\partial p/\partial\epsilon)_n + {n}/{(\epsilon {+} p)}  (\partial p/\partial n)_\epsilon$ is the speed of sound, and $\gamma_s = \coeff43 \eta + \zeta$. Causality demands $0 < c_s^2 < 1$. The first factor (shear waves) gives the causality constraint $\theta_1>\eta$, while the second factor constrains $\delta > T\sigma$. The causality constraints from the last factor are
\begin{align}
  & 0 < v_s^2 \varepsilon_2 - \gamma_s  < \varepsilon_1 \,,\\
  & (1-v_s^2) (\varepsilon_1 - \varepsilon_2) \theta_1 > \gamma_s (\varepsilon_1 + \theta_1) \,.
\end{align}
These constraints can be satisfied for any value of $v_s$ between 0 and 1. Demanding that the modes are stable at ${\bf k}=0$, it is sufficient to require
\begin{align}
  T \left( \frac{\partial p}{\partial\epsilon}\right)_{\!n} \varepsilon_1 + \left( \frac{\partial p}{\partial n}\right)_{\!\epsilon} \varepsilon_3 > 0\,.
\end{align}

\section{Real space analysis of causality}
\label{sec:realspace}
The hydrodynamic equations with the constitutive relations (\ref{eq:cr-2}) are quasi-linear partial differential equations of the form
\begin{equation}
  \left({M}^{\mu\nu}\right)_{AB} \partial_\mu \partial_\nu U^B + \left(\text{lower-derivative terms}\right) = 0 \,,
\end{equation}
where the vector $U$ contains hydrodynamic variables, for example $U^M = (\beta^\mu, \alpha)$. The principal part $M^{\mu\nu}(U)$ is determined by the constitutive relations, and can be read off from the hydrodynamic conservation laws, see ref.~\cite{Hoult:2020eho} for examples. The characteristic velocities of the system may be found by analyzing the roots of the characteristic equation
\begin{equation}\label{eq:char-eq}
\det\left( {M}^{\mu\nu} \xi_\mu \xi_\nu  \right) = 0,
\end{equation}
see e.g.~\cite{Courant-Hilbert}, ch.~VI.
The co-vectors $\xi_\mu$ determined by this equation are normal to the characteristics of the system which must fall within the lightcone. We thus demand that the solutions of \eqref{eq:char-eq} satisfy:
\begin{enumerate}
\item $\xi_0 = \xi_0(\xi_i)$ are real (hyperbolicity), and
\item $|\xi_0| \leq |\vec{\xi}|$ (causality).
\end{enumerate}
In first-order hydrodynamics,  $Q_{AB}\equiv \left({M}^{\mu\nu}\right)_{AB} \xi_\mu \xi_\nu$ is of the form
\begin{equation}
\label{eq:Qmatr}
Q = \begin{bmatrix}
  A u^\nu u_\rho + B \Delta^{\nu}_{\,\,\,\rho} + C u^\nu \xi_\rho + D \xi^\nu u_\rho + E \xi^\nu \xi_\rho & F u^\nu + G \xi^\nu\\
H u_\rho + I \xi_\rho & J
\end{bmatrix}
\end{equation}
In $d+1$ spacetime dimensions, $Q$ is a $(d{+}2)\times(d{+}2)$ matrix. One can derive an explicit expression for the determinant of \eqref{eq:Qmatr}, which is also valid in curved space:
\begin{align}
\label{eq:generic_det}
\det\left(Q\right) &= B^{d-1} \biggl\{ B ( F H - A J) + B (   C J + D J - H G - F I) (\xi{\cdot}u) +  B ( G I - E J ) (\xi{\cdot}u)^2\nonumber\\
&+ \biggl[ G (A I - H C) + F ( H E - I D) + J (C D - A E)\biggr] (\xi{\cdot}\Delta{\cdot}\xi) \biggr\} \,.
\end{align}
This formula facilitates the real-space analysis of causality for relativistic fluids with a conserved global $U(1)$ charge. 
When the principal parts of the hydrodynamic equations from Appendix~\ref{ap::concrete} are inserted into \eqref{eq:generic_det}, one finds the same causality constraints as stated there.

\section{Matching Conditions} \label{ap::matching}
\noindent
An alternative, indirect approach to fixing the zero-modes that arise from the linearized collision operator are so-called ``matching conditions". For example, in order to fix the zero modes in such a way as to arive at the Landau frame, one could impose that the corrections to the energy density are zero, i.e.
\setcounter{equation}{0}
\begin{equation}\label{eq:matchE}
\E-\epsilon = \int_p (p{\cdot}u)^2 \f \left(1\pm\f\right) \phi_p = 0 \,.
\end{equation}
This has the effect of tuning the zero modes such that $\mathcal{E} = \epsilon$. Similarly, one may tune to zero the corrections to the charge density
\begin{equation}\label{eq:matchN}
\N - n = -\int_p (p{\cdot}u) \f \left(1 \pm \f \right) \phi_p = 0\,,
\end{equation}
and the heat current
\begin{equation}\label{eq:matchQ}
\Q^\mu = -\int_p (p{\cdot}u) p_{\perp}^{\mu} \f \left(1 \pm \f \right) \phi_p = 0 \,.
\end{equation}
These matching conditions give only one out of an infinite number of possible hydrodynamic ``frames''. A generalization of these constraints in order to generate causal ``frames'' has been proposed in ref.~\cite{Bemfica:2017wps, Rocha:2021lze}, reading
\begin{subequations}\label{eq:matchcons}
\begin{align}
\int_p (p{\cdot}u)^{r} \f \left(1 \pm \f \right) \phi_p &= 0,\\
\int_p (p{\cdot}u)^{s} \f \left(1 \pm \f \right) \phi_p &= 0,\\
\int_p (p{\cdot}u)^{t} p_{\perp}^{\mu} \f \left(1 \pm \f \right) \phi_p &= 0,
\end{align}
\end{subequations}
where $r \neq s$, and $r,s,t$ are non-negative integers. While the physical meaning of setting to zero the quantities that do not appear in the constitutive relations is not immediately clear, the real question is how the matching conditions \eqref{eq:matchcons} are related to the hydrodynamic field redefinitions which give rise to causal ``frames''. 

The correction to the equilibrium distribution function is given by eq.~\eqref{eq:ff-22},
\begin{equation}
\phi_p = a(x) + b_\mu(x) {p^\mu} + \Phi_p(x) \,,
\end{equation}
where $\Phi_p$ is the inhomogeneous part of the solution, and $a$, $b_\mu$ are the space- and time-dependent parts of the zero modes. 
Substituting this $\phi_p$ into \eqref{eq:matchcons}, we find%
\begin{subequations}
\label{eq:abu}
\begin{align}
   a &= \frac{ \braket{(p{\cdot}u)^r \Phi_p} \braket{(p{\cdot}u)^{s+1}} - \braket{(p{\cdot}u)^{r+1}} \braket{(p{\cdot}u)^s \Phi_p} }{ \braket{(p{\cdot}u)^{r+1}} \braket{(p{\cdot}u)^s} - \braket{(p{\cdot}u)^r} \braket{(p{\cdot}u)^{s+1}} } \,,\\
   b{\cdot}u &=  \frac{\braket{(p{\cdot}u)^r \Phi_p} \braket{(p{\cdot}u)^{s}} - \braket{(p{\cdot}u)^{r}} \braket{(p{\cdot}u)^s \Phi_p} }{\braket{(p{\cdot}u)^{r+1}} \braket{(p{\cdot}u)^s}  - \braket{(p{\cdot}u)^r}  \braket{(p{\cdot}u)^{s+1}} }  \,, \\
   b_\nu^\perp & = -\frac{\langle (p{\cdot}u)^t p^\mu_\perp \Phi_p \rangle }{ \frac{1}{d} \langle (p{\cdot}u)^t p_\perp^2 \rangle } \,,
\end{align}
\end{subequations}
where $\braket{...} = \int_p \left(...\right) \f(1 \pm \f)$. 
Clearly, one needs to know the inhomogeneous solution $\Phi_p$ in order to relate $a$ and $b_\mu$ to $r$, $s$, and $t$. The form of $\Phi_p$ depends on how one chooses to impose the perfect-fluid constraint. Let's say we choose to eliminate the time derivatives as in eq.~\eqref{eq:Phi-1},
\begin{align}
\label{eq:Phi-1a}
  (1\pm \f_p) \Phi_p = K_\eta p^\mu p^\nu \sigma_{\mu\nu}[{u}] + K_\zeta \partial{\cdot}{u} + K_\alpha p^\mu \partial_\mu^\perp {\alpha}\,.
\end{align}
For massless particles $K_\zeta = 0$, and \eqref{eq:abu} implies that $a=b{\cdot}u=0$. In other words, once the perfect-fluid constraint has been imposed when finding the inhomogeneous solution $\Phi_p$, hydrodynamic field redefinitions can not generate causal frames because the latter require non-zero transport parameters in the scalar sector. Put differently, conditions \eqref{eq:abu} imply that once the frame-invariants $f_i$ vanish, the transport parameters $\varepsilon_i, \pi_i, \nu_i$ must vanish as well, which is inconsistent with causal frames in first-order conformal hydrodynamics.

Similarly, for massive particles with $K_\zeta \neq0$, eqs.~\eqref{eq:Phi-1a} and \eqref{eq:abu} imply 
\begin{subequations}
\begin{align}
  a &= (0) \dot T  + \left( \textrm{non-zero}\right) \partial{\cdot}u + (0) \dot{\alpha}\,,\\
  b{\cdot}u &= (0) \dot T  + \left( \textrm{non-zero}\right) \partial{\cdot}u + (0) \dot{\alpha}\,.
\end{align}
\end{subequations}
Again, we see that by fixing the zero modes via matching conditions \eqref{eq:matchcons}, one is unable to generate a suitable mix of time- and space-derivatives required for hyperbolicity and causality. 

The only way to generate a causal frame via the matching conditions \eqref{eq:matchcons} would be if $\Phi_p$ contained independent functions multiplying $\dot{T}/T$, $\partial{\cdot}u$, $\dot{\alpha}$ in the scalar sector. However, as the source term of the linearized Boltzmann equation must obey the perfect-fluid constraints, only one of these three functions is allowed in the inhomogeneous solution.
The issue may be alleviated by taking moments of the full Boltzmann equation, as was done in \cite{Rocha:2021lze} to study non-hydrodynamic contributions.

\bibliographystyle{JHEP}
\bibliography{hydro-general-biblio}
\end{document}